\documentstyle[11pt]{article}

\def\be {\begin{eqnarray}}
\def\ee {\end{eqnarray}}
\def\beq {\begin{equation}}
\def\eeq {\end{equation}}
\def\Tr{{\rm Tr}}

\def\pl {Phys. Lett.}

\def\bitem {\begin{itemize}}
\def\eitem {\end{itemize}}
\def\ben {\begin{enumerate}}
\def\een {\end{enumerate}}

\begin{document}
\hfill {\it \today}

\vskip 0.4in
\begin{center}
{\large\bf VANISHING OF NONABELIAN BERRY PHASES}\\
{\large\bf IN HEAVY SOLITONIC BARYONS}
\end{center}
\vskip 0.6in
\centerline{Hyun Kyu Lee\footnote{Permanent address: Department of Physics,
Hanyang University, Seoul 133-791, Korea. Supported in part by the KOSEF
under Grant No.91-08-00-04 and by Ministry of Education (BSRI-92-231)}
and Mannque Rho}

\vskip 0.4in

\centerline{\it Service de Physique Th\'{e}orique}
\centerline{\it C.E. Saclay}
\centerline{\it F-91191 Gif-sur-Yvette, France}

\vskip .7in
\centerline{\bf ABSTRACT}
\begin{quotation}
{\noindent
In this article, we show -- using a reasoning applicable to both the
excitation
spectrum in the chiral bag model and the hyperfine structure of diatomic
molecules -- that
the generic form of a nonabelian Berry potential appears in heavy quark
effective theory (HQET) and that the Berry potential vanishes
for the soliton-heavy meson bound state in the heavy-quark
limit. The vanishing of the Berry potential in HQET is shown to be
related to the restoration of heavy-quark symmetry in
infinite heavy-quark-mass limit in close analogy to diatomic molecules
at infinite internuclear separation.}
\end{quotation}

\newpage
\section{Introduction}
\indent

In a series of recent papers\cite{lnrz0}, it has been
shown that Berry structures\cite{berry}
can appear naturally in hadron physics in the framework of the chiral
bag model.
More recently, the Berry structure was also identified in the
spectrum of heavy-quark baryons\cite{mopr,nrz2}, providing an
understanding of heavy-quark symmetry \cite{hqe} in terms of a
``symmetry restoration." In particular, it was shown in \cite{mopr,nrz2}
that when a baryon with a heavy quark is described in terms of the binding
of a heavy meson with an $SU(2)$ soliton \`{a} la Callan and Klebanov (CK)
\cite{ck}, as the mass of the heavy quark increases, the Wess-Zumino term
binding of Callan and Klebanov ceases to be operative as the Wess-Zumino
term disappears but the binding survives with, however,
the hyperfine coefficient $c$ that gives the splitting between, say, $\Sigma$
and $\Sigma^*$
vanishing as $1/m_H$ where $m_H$ is the heavy-meson mass.

In this paper, we present a simple argument that explains how a Berry
structure\cite{berry} emerges in heavy quark effective theory\cite{hqe}
and how the Berry potential vanishes in the heavy-quark limit using the
quantum mechanical binding mechanism of Manohar et al.\cite{manohar}.
We will use a reasoning completely analogous to that employed for the
chiral bag\cite{lnrz0} on the one hand and that used for the restoration
of electronic rotational invariance in diatomic molecules \cite{zg} on the
other. We will see that this provides yet another
demonstration in support of previous arguments of Refs.\cite{mopr,nrz2}
that a generic Berry structure is playing a key role in the heavy-baryon
structure.
\section{Diatomic Molecules}
\indent

In order to better understand the complex situation of strongly interacting
systems that we are interested in, we first discuss a generic case of
quantum mechanical system, namely the diatomic molecule. To do this,
we begin by recalling here a well-known fact \cite{jakman} that the
conservation laws of a system coupled to
a symmetric background gauge field persist in modified forms. The simplest
example is the angular momentum of a system coupled to a Dirac ($U(1)$)
magnetic monopole. Here the conserved angular momentum is modified to a
sum of the usual mechanical angular momentum and a contribution from
the magnetic field\cite{yang}. A more
interesting phenomenon related to this observation is ``spin-isospin"
transmutation \cite{spin}.
\subsection{Non-abelian Berry Potential}
\indent

As shown by Berry \cite{berry}, in a quantum system, induced gauge fields
naturally appear in the space
of slow variables when the fast variables are integrated out. They are
referred to in the literature as Berry potentials.  The Schr\"{o}dinger
equation resulting after fast variables are integrated out
is given by the following form,
\be
-\frac{1}{2m}(\vec{\nabla}_{R}  - i\vec{{\bf A}})^2 {\bf \Psi} =
i \frac{\partial {\bf \Psi}}{\partial t}\label{schrod}
\ee
where $\vec{{\bf A}}$ is defined by
\be
\vec{{\bf A}}_{a,b} = i \langle a,\vec{R}|\vec{\nabla}|b,\vec{R}\rangle.
\label{amn}
\ee
$|a,\vec{R}\rangle$ is a `snap-shot' eigenstate for a given {\it slow
variable} $\vec{R}$, which
is related to the reference state $|a\rangle$ by a unitary  operator
$U(\vec{R})$ such that
\be
|a,\vec{R}\rangle = U(\vec{R})  |a\rangle.\label{eqU}
\ee
Equation (\ref{schrod}) is a matrix equation where ${\bf \Psi}$ is a
column vector defined in a vector space described by $|a\rangle$.

It is convenient to introduce, following \cite{casal}, a
Grassmann variable $\theta_a$ for each $|a\rangle$ (say, an electronic
state) and rewrite the Schr\"{o}dinger equation Eq.(\ref{schrod}) as
\be
-\frac{1}{2m}(\vec{\nabla}_{R}  - i\vec{{\bf A}}
(\theta,\theta^{\dagger},\vec{R}))^2 {\bf \psi}(\theta, \vec{R}) =
i \frac{\partial {\bf \psi}(\theta, \vec{R})}{\partial t}.\label{gschrod}
\ee
In the above equation, internal degrees of freedom are considered to be
dynamical degrees of freedom treated classically in the form of anticommuting
coordinates.    Equation (\ref{gschrod}) can be obtained by quantizing
the system described by the following Lagrangian
\be
{\cal L}=\frac{1}{2}m\dot{\vec{R}}^2 + i\theta^{\dagger}_a(\frac{\partial}
{\partial t}-i\vec{A}^{\alpha}T^{\alpha}_{ab}\cdot\dot{\vec{R}})\theta_b
\label{glagran}
\ee
where ${\bf T}^{\alpha}$ is a matrix representation in the vector space of
$|a\rangle$'s for a generator ${\bf {\cal T}}^{\alpha}$ of $U(\vec{R})$,
\be
[{\bf T}^{\alpha},{\bf T}^{\beta}]=i f^{\alpha\beta\gamma}{\bf T}^{\gamma}.
\label{talgeb}
\ee
Following the standard quantization procedure \cite{casal,diracq}, we obtain
the following commutation relations,
\be
[R_i,p_j]=i\delta_{ij}, \ \ \ \ \ \{\theta_a,\theta_b^{\dagger}\}=i\delta_{ab}.
\label{quantiz}
\ee
It is then straightforward to obtain the Hamiltonian
\be
H=\frac{1}{2m}(\vec{p}-\vec{{\bf A}})^2\label{thetah}
\ee
where
\be
\vec{{\bf A}} &=& \vec{A}^{\alpha}I^{\alpha},\nonumber \\
   I^{\alpha} &=& -i\theta^{\dagger}_a T^{\alpha}_{ab}\theta_b.\label{ialpha}
\ee
Using the commutation relations, it can be verified that
\be
[I^{\alpha},I^{\beta}]=if^{\alpha\beta\gamma}I^{\gamma}.
\ee
The Schr\"{o}dinger equation
\be
H\psi=i\frac{\partial \psi}{\partial t},
\ee
with Eq.(\ref{thetah}) leads to Eq.(\ref{gschrod}).
It is clear that the Lagrangian, Eq.(\ref{glagran}), is
invariant under the gauge transformation
\be
\vec{A}^{\alpha} &\rightarrow& \vec{A}^{\alpha} + f^{\alpha\beta\gamma}
\Lambda^{\beta}\vec{A}^{\gamma} - \vec{\nabla}\Lambda^{\alpha}\label{gaugea},
\\
\theta_a &\rightarrow& \theta_a -i\Lambda^{\alpha}T^{\alpha}_{ab}
\theta_b.\label{gaugetr}
\ee
It should be noted that  Eq.(\ref{gaugetr})
corresponds to the gauge transformation on $|a\rangle$. This makes gauge
invariance manifest in the Lagrangian. We should also point out
that the Lagrangian (\ref{glagran}) resembles closely {\it both} the Lagrangian
obtained in \cite{lnrz0} for the chiral bag model of baryon structure
when the sea quarks are integrated out {\it and} the Lagrangian that emerges
in heavy-quark effective theory discussed below.

\subsection{Conserved angular momentum}
\indent

Consider a particle coupled to an external gauge field of 't Hooft
 -Polyakov monopole\cite{monotp} with a coupling constant $g$.
The asymptotic form of the magnetic field is given by
\be
\vec{\bf B} = -\frac{\hat{r}(\hat{r}\cdot{\bf T})}{gr^2}\label{bmag}
\ee
which is obtained from the gauge field $\vec{{\bf A}}$
\be
A^{\alpha}_i &=& \epsilon_{\alpha i j} \frac{r_j}{gr^2},
\label{hooft}\\
\vec{{\bf B}} &=& \vec{\nabla} \times \vec{{\bf A}} -ig[\vec{{\bf A}},
\vec{{\bf A}}].\label{twoform}
\ee
Using the convention described in the previous section, the Hamiltonian of
a particle coupled to a 't Hooft-Polyakov monopole can be written as

\be
H &=& \frac{1}{2m}(\vec{p}-\vec{\bf A})^2\nonumber\\
  &=& \frac{1}{2m}\vec{{\bf D}}\cdot \vec{{\bf D}}\label{dhamil}
\ee
where $\vec{{\bf D}}=\vec{p}-\vec{\bf A}$. Here and in what follows
 we put $g=1$ for close analogy with Eq.(\ref{thetah})
Obviously the mechanical angular momentum $\vec{L}_m$ of a particle
\be
\vec{L}_m=m\vec{r}\times\dot{\vec{r}}=\vec{r}\times\vec{{\bf D}}
\ee
does not satisfy the $SU(2)$ algebra after canonical quantization in
Eq.(\ref{quantiz})
and moreover cannot be a symmetric operator that commutes with the
Hamiltonian. Of course
the conventional angular momentum, $\vec{L}_o =\vec{r}\times\vec{p}$,
satisfies the usual angular momentum commutation rule, but
it does not commute with the
Hamiltonian and hence cannot be a conserved angular momentum of the
system.  This shows that the construction of a conserved angular
momentum of a system coupled to a topologically nontrivial gauge field is not
a trivial matter.

To construct the conserved angular momentum, we have to modify $\vec{L}_m$ to
\be
\vec{L} = \vec{L}_m + \vec{{\bf Q}},\label{rxd}
\ee
with $ \vec{{\bf Q}}=\vec{Q}^{\alpha}I^{\alpha}$ to be determined.
The methods to determine $\vec{{\bf Q}}$ have been discussed in the
literature \cite{jakman,yang}. Here we adopt a rather straightforward
method.
The first condition required for $\vec{{\bf Q}}$ is the consistency condition
that $\vec{L}$ satisfy the $SU(2)$ algebra
\be
[L_i,L_j]=i\epsilon_{ijk}L_k.\label{lll}
\ee
This leads to an equation for $\vec{{\bf Q}}$,
\be
\vec{r}(\vec{r}\cdot\vec{{\bf B}}) + \vec{r}\vec{{\cal D}}\cdot\vec{{\bf Q}}
-\vec{{\cal D}}(\vec{r}\cdot\vec{{\bf Q}})=0\label{c1}
\ee
where
\be
\vec{{\cal D}}=\vec{\nabla} - i[\vec{{\bf A}},\ \ \ ].\label{covd}
\ee
The second condition is obtained by requiring that $\vec{L}$ commute with H,
\be
[\vec{L},H]=0.\label{c2}
\ee
Equation (\ref{c2}) can be replaced by a stronger condition
\be
 [L_i, {\bf D}_j]=i\epsilon_{ijk}{\bf D}_k,\label{c3}
\ee
which leads to
\be
{\cal D}_i{\bf Q}_j +\delta_{ij}\vec{r}\cdot\vec{{\bf B}} -
r_i{\bf B}_j=0. \label{c31}
\ee
It is obvious that $\vec{L}$ satisfying Eq.(\ref{c3}) or (\ref{c31}) commutes
with the Hamiltonian Eq.(\ref{dhamil}).
Equation (\ref{c31}) is just the condition for ``spherically symmetric
potential" discussed by Jackiw\cite{jackiw}. Here we can verify it in a more
straightforward way
using Eq.(\ref{c3}). Moreover the meaning of spherical symmetry becomes clear
from Eq.(\ref{c2}).

In the case of  the  't Hooft-Polyakov monopole, Eqs.(\ref{bmag})
and (\ref{hooft}), it can be shown that
\be
\vec{{\bf Q}} = \hat{r}(\hat{r}\cdot{\bf I})\label{phitm}
\ee
satisfies Eqs. (\ref{c1}) and (\ref{c31}). After inserting Eq.(\ref{phitm})
into Eq.(\ref{rxd}), we get
\be
\vec{L} &=& \vec{L}_m + \hat{r}(\hat{r}\cdot{\bf I})\label{ltm0}\\
        &=& \vec{r} \times \vec{p} + \vec{I},\label{ltm}
\ee
where
\be
I_i=\delta_{i\alpha}I^{\alpha}.\label{ii}
\ee
Equation (\ref{ltm}) with (\ref{ii}) shows clearly how the isospin-spin
transmutation takes place in a
system where a particle is coupled to a nonabelian monopole.

The same analysis can be applied to the abelian $U(1)$ monopole
just by replacing $\hat{r}\cdot\vec{I}$ by -1 in Eqs.(\ref{phitm}) and
(\ref{ltm0}): We are considering a Dirac monopole with $e=g=1$. Then
\be
\vec{Q} &=& \hat{r},\label{phidm}\\
\vec{L}    &=& m\vec{r}\times\dot{\vec{r}} - \hat{r}.\label{ldm0}
\ee
One can rewrite Eq.(\ref{ldm0}) in a more familiar form seen
in the literature
\be
\vec{L} = \vec{r}\times\vec{p} - \vec{\Sigma},\label{ldm}
\ee
where
\be
\vec{\Sigma} =\left(\frac{(1-\cos\theta)}{\sin\theta}\cos\phi,\,
\frac{(1-\cos\theta)}{\sin\theta}\sin\phi,\, 1\right).
\ee

\subsection{Rotational symmetry of nonabelian Berry potential}
\indent

So far our consideration has been rather general. Let us now
focus on  conserved angular momentum in a diatomic
molecule in which a Berry potential couples to the dynamics of slow degrees of
freedom, corresponding to the nuclear coordinate $\vec{R}$. This
system has been studied by Zygelman\cite{zg}.

The Berry potential is defined on the space spanned by the electronic
states $\pi(|\Lambda|=1)$ and $\Sigma(\Lambda=0)$, where $\Lambda$'s are
eigenvalues of the third component of the orbital angular momentum of
the electronic
states.  The electronic states responding to the slow rotation $U(\vec{R})$
of $\vec{R}$ defined by
\be
U(\vec{R})={\rm exp}(-i \phi L_z){\rm exp}(i \theta L_y){\rm exp}(i \phi L_z),
\label{uzg}
\ee
induce a Berry potential of the form
\be
\vec{{\bf A}} &=& i\langle \Lambda_a|U(\vec{R})\vec{\nabla}U(\vec{R})^{\dagger}
|\Lambda_b\rangle\label{zga0}\\
              &=& \frac{{\bf A}_{\theta}}{R} \hat{{\bf \theta}} +
\frac{{\bf A}_{\phi}}{R \sin\theta}\hat{{\bf \phi}},
\ee
where
\be
{\bf A}_{\theta}&=&\kappa(R)({\bf T}_y \cos\phi - {\bf T}_x \sin\phi),
\nonumber\\
{\bf A}_{\phi}&=& {\bf T}_z(\cos\theta - 1) - \kappa(R)\sin\theta
({\bf T}_x \cos\phi + {\bf T}_y \sin\phi). \label{aphi}
\ee
Here $\vec{{\bf T}}'$s are spin-1 representations of
the orbital angular momentum
$\vec{L}$ and $\kappa$ measures the transition amplitude between the
$\Sigma$ and $\pi$ states
\be
\kappa(R)=\frac{1}{\sqrt{2}}|\langle\Sigma|L_x-iL_y|\pi\rangle|.\label{kappa}
\ee
The nonvanishing field strength tensor is given by
\be
\vec{{\bf B}}=\frac{F_{\theta\phi}}{R^2 \sin\theta}=-\frac{(1-\kappa^2)}
{R^2}T_z\hat{R}.\label{zgb0}
\ee
Following the procedure described in section 2.1, we introduce a Grassmann
variable for each electronic state. Replacing ${\bf T}$ by {\bf I} defined
in Eq.(\ref{ialpha}) and quantizing the corresponding Lagrangian, we obtain
the Hamiltonian
\be
H=\frac{1}{2\mu} (\vec{p} - \vec{{\bf A}})^2 \label{hzg0},
\ee
where $\vec{{\bf A}} = \vec{A}^{\alpha} I^{\alpha}$ and $\mu$ is the reduced
mass.
The presence of the constant $\kappa$ -- which is not quantized --
in the
Berry potential is a generic feature of nontrivial nonabelian Berry potentials
as can be seen in many examples \cite{gp,lnrz0}.

To find a solution of Eq.(\ref{c1}) and
Eq.(\ref{c31}), it is better to look into the Hamiltonian in detail.
Exploiting the gauge invariance, the Hamiltonian can be rewritten in the  most
symmetric form. This can be done by subtracting a trivial (or pure gauge) part
out of
the Berry potential, which is equivalent to choosing a new gauge such that
\be
\vec{{\bf A}}'&=& V^{\dagger}\vec{{\bf A}}V + iV^{\dagger}\vec{\nabla}V
\label{zgas}\\
        {\bf F}' &=& V^{\dagger}{\bf F}V\label{zgbs}
\ee
where $V$ is an inverse operation of $U$ in Eq.(\ref{uzg}), {\it i.e.},
$V = U^{\dagger}$.  Then
\be
{\bf A}_{\theta}' &=& (1-\kappa)(I_x \sin\phi-I_y \cos\phi),\nonumber\\
{\bf A}_{\phi}' &=& (1-\kappa)\{-I_z \sin^2 \theta +
\cos\theta \sin\theta (I_x \cos\phi + I_y \sin\phi )\},\label{zgasf}
\ee
or more compactly
$$\vec{{\bf A}}' = (1-\kappa)\frac{\hat{R} \times \vec{I}}{R^2},$$
and
\be
\vec{{\bf B}}'=-(1-\kappa^2)\frac{\hat{R}(\hat{R}\cdot{\bf I})}
{R^2}.\label{zgbsf}
\ee
A remarkable feature of the above Berry potential is that it has the same
structure as the 't Hooft-Polyakov monopole, Eq.(\ref{bmag}) and
Eq.(\ref{hooft}), but with different constant factors,
$(1-\kappa)$ for vector potential and $(1-\kappa^2)$ for magnetic field.
Because of these two different factors, one cannot simply
take Eq.(\ref{phitm}) as a solution of (\ref{c31}) for the case of
nonabelian Berry potentials.

Using the following identities derived from Eq. (\ref{zgasf}),
\be
\vec{R}\cdot\vec{{\bf A}}' &=& 0,\label{da}\\
\vec{R} \times \vec{{\bf A}}' &=& -(1-\kappa)\{\vec{I} -
(\vec{I}\cdot\hat{R})\hat{R}\},\label{ca}
\ee
the Hamiltonian, Eq.(\ref{hzg0}), can be written as
\be
H = -\frac{1}{2\mu R^2}\frac{\partial}{\partial R}R^2\frac{\partial}
{\partial R}
+ \frac{1}{2\mu R^2}(\vec{L}_o + (1-\kappa)\vec{I})^2
- \frac{1}{2\mu R^2}(1-\kappa)^2(\vec{I}\cdot\hat{R})^2.\label{hzg1}
\ee
One sees from this Hamiltonian that the factor $(1-\kappa)$ controls
hyperfine splitting. For this reason we call it ``hyperfine
coefficient." The corresponding quantity in heavy baryons is denoted
$c$ in the later section.

Now one can show that the conserved angular momentum $\vec{L}$ is
\be
\vec{L} &=&  \vec{L}_o + \vec{I},\label{zglf}\\
        &=&  \mu\vec{R}\times\dot{\vec{R}} + \vec{{\bf Q}},\label{zglmf}
\ee
with
\be
\vec{{\bf Q}} = \kappa \vec{I} + (1-\kappa)\hat{R}(\hat{R}\cdot\vec{I}).
\label{zgqf}
\ee
Hence, in terms of the conserved angular momentum $\vec{L}$,
the Hamiltonian becomes
\be
H = -\frac{1}{2\mu R^2}\frac{\partial}{\partial R}R^2\frac{\partial}
{\partial R}
+ \frac{1}{2\mu R^2}(\vec{L} - \kappa\vec{I})^2
- \frac{1}{2\mu R^2}(1-\kappa)^2\label{hzg2}
\ee
where  $(\vec{I}\cdot\hat{R})^2 = 1$ has been used.

It is interesting to see what happens in the two extreme cases of $\kappa = 0$
and $1$. For $\kappa=0$, the
$\Sigma$ and $\pi$ states are completely decoupled and only the
$U(1)$ monopole field can be developed on the $\pi$
states\cite{moody}. Equation (\ref{zgqf}) becomes identical
to Eq.(\ref{phidm}) as
$\kappa$ goes to zero and the Hamiltonian can be written as
\be
H = -\frac{1}{2\mu R^2}\frac{\partial}{\partial R}R^2\frac{\partial}
{\partial R}
+ \frac{1}{2\mu R^2}(\vec{L}\cdot\vec{L} - 1)\label{hzgu1}
\ee
which is a generic form for a system coupled to an $U(1)$ monopole field.
Physically this corresponds to small internuclear distance at which
the $\Sigma$ and $\pi$ states decouple.

For $\kappa = 1$, the degenerate $\Sigma$
 and $\pi$ states form a representation of the rotation group and hence the
Berry potential (and its field tensor) vanishes or becomes a pure gauge.
Then $\vec{{\bf Q}} = \vec{I}$ and $\vec{L} = \mu\vec{R}\times\dot{\vec{R}} +
\vec{I} $.  Now $\vec{I}$ can be understood as the angular momentum of the
electronic system which is
decoupled from the spectrum. One can also understand this as the restoration
of rotational symmetry in the electronic system. Physically $\kappa \rightarrow
1$ as $R\rightarrow \infty$.
In the next section, we shall show that the same situation occurs in
heavy-quark effective theory where the restoration of the heavy-quark
symmetry  for $m_Q\rightarrow \infty$ is manifested by a vanishing
Berry potential.

\section{Heavy-Quark Baryons}
\indent

We shall now show that the same generic form of Berry potentials emerges
in the spectrum of heavy-quark baryons (containing heavy quarks c, b etc)
and that the heavy-quark symmetry discovered in QCD can be identified
with the vanishing of nonabelian Berry potentials in the symmetry limit.

\subsection{Emergence of Berry potentials}
\indent

Consider the effective Lagrangian with chiral symmetry for light quarks
and heavy quark symmetry for heavy quarks \cite{manohar}
\be
{\cal L} &=& -i Tr \overline{H}_a v^\mu \partial_\mu H_a
 + i Tr \overline{H}_a H_b v^\mu \left(\Sigma^{\dagger}\partial_{\mu}
\Sigma \right)_{ba} \nonumber \\
& & + ig Tr \overline{H}_a H_b \gamma^\mu \gamma^5
\left(\Sigma^{\dagger}\partial_{\mu}\Sigma \right)_{ba} +
\frac{F_{\pi}^2}{16}tr\left(\partial_{\mu}\Sigma \partial^{\mu}
\Sigma^{\dagger}\right) + \cdots \ \   \label{hq}
\ee
which we rewrite in the rest frame of the heavy quark, $v^{\mu} = (1,0)$,
\be
{\cal L} &=& -i Tr \overline{H}_a \partial_t H_a
 + i Tr \overline{H}_a H_b  \left(\Sigma^{\dagger}\partial_t
\Sigma \right)_{ba} \nonumber \\
& & + ig Tr \overline{H}_a H_b \gamma^\mu \gamma^5
\left(\Sigma^{\dagger}\partial_{\mu}\Sigma \right)_{ba} +
\frac{F_{\pi}^2}{16}Tr\left(\partial_{\mu}\Sigma \partial^{\mu}
\Sigma^{\dagger}\right) + \cdots \ \ .  \label{hqt}
\ee
Here $\Sigma$ is the usual chiral $SU(2)$ field, $H$ the heavy-meson field
with the quark configuration $Q\bar{q}$ (where $Q$ denotes a heavy quark and
$q$ a light quark)
consisting of the pseudoscalar $P$ and the vector $P^*_\mu$ and $F_\pi$
the pion decay constant $\approx 186$ MeV. Throughout we shall follow the
notation of \cite{manohar}.

To see how Berry potentials arise, we must identify ``fast"
and ``slow" variables (or degrees of freedom)
in the theory.  We take the light anti-quark of a heavy meson as a ``fast"
variable
to be ``integrated out" in the presence of a slowly rotating
soliton background constructed in the light meson sector which represents
a ``slow" variable, while the heavy quark
is ``integrated in" as a spectator. This can be visualized by assuming
that the soliton-heavy meson bound
system is composed of a heavy quark $Q$ sitting at the origin and a light
antiquark in the heavy meson $H$, $\bar{q}_H$, moving in the background of
a slowly rotating soliton.
The heavy quark then is effectively decoupled from the soliton, so it
is a spectator in the limit $m_Q\rightarrow \infty$.
Therefore, as far as the rotation motion of the soliton is concerned,
the effective degree of freedom
is the light antiquark in the heavy meson that couples to the soliton. Thus
one can take the ``fast" variable in this system to be the light antiquark
$\bar{q}_H$ instead of the whole heavy meson itself as in the CK picture
used in \cite{mopr,nrz2}. It is a ``fast" variable in the sense that the
excitation due to the motion of the slow variable (that is to say,
the adiabatic rotation
of the soliton) is of order $1/N_c$ while the energy splitting of the
soliton-antiquark system is of order $N^0_c$ which depends on the grand spin
$K$ as explained below. This means that the adiabatic theorem -- which
states that the fast degree
of freedom continues to remain on its snapshot
eigenstate up to a phase -- can be used to obtain a Berry potential defined in
the space of the slow variable. As we shall show below, there is a
close analogy to the chiral bag case \cite{lnrz0} where a quark (fast
variable) moves in the background field of a rotating soliton (slow variable).
We shall exploit this analogy in simplifying our argument.

The next step in our argument
is to find the spectrum of the ``fast" variable -- $H$ -- in the solitonic
background {\it before} rotating the soliton. The relevant part of the
Lagrangian involving the $H$ field can be obtained by
replacing $\Sigma$ by the time-independent soliton (hedgehog) field
$\Sigma_o$ in Eq.(\ref{hqt}),
\be
{\cal L}_H &=& -i Tr \overline{H}_a \partial_t H_a
 + ig Tr \overline{H}_a H_b \gamma^i \gamma^5
\left(\Sigma_o^{\dagger}\partial_i\Sigma_o \right)_{ba} + \cdots\ \ .\label{lh}
\ee
The second term of (\ref{lh}) is an interaction term which can be written
in terms of the isospin $I_H$ and the angular momentum $S_{l}$ of the light
degree of freedom, $\bar{q}_H$, in the heavy meson\cite{manohar,foot1}
\be
H_I = V_I (\Sigma_o) \left(\vec{I}_H \cdot \vec{S}_{l} \right). \label{hin}
\ee
Hence the energy eigenstates are classified by the $K$-spin of {\em
$\bar{q}_H$},
$K=I_H
+ S_{l}$ (recall that the hedgehog $K$-spin is zero)
with an energy splitting $\Delta E = V_I^K$, which is of order $N_c^0$.
The detailed calculations of $V_I (\Sigma_o)$ will be discussed later.
What we wish to point out here is that
the interaction Hamiltonian does not ``see" the orbital angular momentum of
$\bar{q}_H$. Therefore the orbital angular momentum is separately conserved.

We now rotate the chiral field with the slow variable $S(t)$
\be
\Sigma = S(t) \Sigma_o S^{\dagger}(t)\label{ar}
\ee
in Eq.(\ref{hqt}) but leave unrotated the heavy-meson field $H$ as proposed by
Manohar et al.\cite{manohar}. This contrasts with the quantization procedure of
\cite{mopr,nrz2} where the $H$ field is also rotated. The difference
is just a matter
of choosing frames, the former corresponding to the rotating inertial frame
and the latter to the heavy-meson rest frame.
Now with Eq. (\ref{ar}) and Eq.(\ref{hqt}), the Lagrangian can be
written in the form
\be
{\cal L}_H &=& -i Tr \overline{H} \partial_t H \nonumber \\
      & & + i Tr \overline{H}_a H_b  \left(\Sigma^{\dagger}\partial_t
          \Sigma \right)_{ba} \nonumber \\
   & & + ig Tr \overline{H}_a H_b \gamma^0 \gamma^5
\left(\Sigma^{\dagger}\partial_t \Sigma \right)_{ba}\nonumber \\
   & & + ig Tr \overline{H}_a H_b \gamma^i \gamma^5
\left(S^{\dagger}(\Sigma_o^{\dagger}\partial_i\Sigma_o)S \right)_{ba}.
\label{li}
\ee
If we assume that the bound state is formed at
the origin of the soliton as discussed by Guralnik et al.\cite{manohar},
then the second and third terms of Eq. (\ref{li})
vanish. This can be seen by the fact that
\be
\Sigma^{\dagger}(0)\partial_t\Sigma(0) &=& S\Sigma^{\dagger}_0(0)S^{\dagger}
\partial_t S \Sigma_0(0) S^{\dagger}+ \partial_t S^{\dagger}S\nonumber\\
&=&  S^{\dagger}\partial_t S +\partial_t S^{\dagger}S =0\label{r0}
\ee
where the value of $\Sigma_0$ at $r=0$, $\Sigma_0(0)=1$, has been used.
Then from the reduced Lagrangian
\be
{\cal L}_H &=& -i Tr \overline{H} \partial_t H \nonumber \\
    & & + ig Tr \overline{H}_a H_b \gamma^i \gamma^5
\left(S^{\dagger}(\Sigma_o^{\dagger}\partial_i\Sigma_o)S \right)_{ba},
\label{li1}
\ee
the interaction Hamiltonian can be obtained as
\be
H_I(t) =S H_I S^{\dagger}.\label{hint}
\ee
It seems natural to take $S(t)|K\rangle$ to be the snapshot eigenstate
at t, $\Psi_K (t)$, where $|K\rangle$ is an energy eigenstate of the unrotating
hedgehog soliton ({\it i.e.}, S(0)=1).  The energy eigenstate of $H_I(t)$
evolves accordingly along the path determined
by the snapshot eigenstate $\Psi_K(t)$ up to a phase, generating
a Berry potential in Born-Oppenheimer approximation.
The Berry potential can then be calculated in a standard way
\be
{\cal A} &=& i\langle K S^{\dagger}|\partial |S K\rangle\nonumber\\
         &=& i\langle K|( S^{\dagger}\partial S)|K\rangle.\label{v1}
\ee
The key feature of this procedure is that the Berry potential can
be calculated in a reference state $|K\rangle$, which measures in fact how
the $K$-states defined at t=0 get mixed during the adiabatic rotation.
This also shows clearly that
$\cal A$ is an induced gauge potential coupled to a slow variable
which is the rotation of the soliton.

An equivalent but more instructive way to exhibit the Berry structure is
to redefine the $H$ field in such a way that the rotation matrix $S$
acts on the heavy meson field rather than on the soliton.
Let
\be
H^\prime = HS^{\dagger}, \ \ \, \overline{H}^\prime = S\overline{H}.
\ee
Now the action -- which is a quantum mechanical Lagrangian for the slow
variable
$S(t)$ -- becomes (dropping the prime of $H^\prime$)
\be
L &=&\int d^3x \left[ -i \Tr \overline{H} \partial_t H
     -i \Tr H (\partial_t S^{\dagger} S) \overline{H}
     + ig \Tr \overline{H}_a H_b \gamma^i \gamma^5
\left(\Sigma_o^{\dagger}\partial_i\Sigma_o \right)_{ba}\right]\nonumber \\
   & & + \frac{{\cal I}}{4}\Tr(S^{\dagger} \partial_t  S)^2 + \cdots \ \
\label{la}
\ee
where we have now restored the kinetic energy term for the slow degree of
freedom with ${\cal I}$ the moment of inertia of the rotating soliton.
The third term of (\ref{la}) involves no time derivative and hence
is a potential which we denoted above as $(-H_I)$ on which we will have more
to say below.  The second term can be identified as the Berry potential
which gives the same
${\cal A}$ as in Eq. (\ref{v1}) when projected onto the $K$ state.
That this identification is sensible can be seen by making an
analogy to the chiral bag. In the case of the chiral bag discussed in
\cite{lnrz0}, the adiabatic change of the skyrmion, $S(t)$,
is incorporated in the following Lagrangian,
\be
L^{CB} = \int_V d^3x \left[\overline{\psi}
i\gamma^{\mu}\partial_{\mu}\psi
-\frac 12 \Delta_s \,\,
\overline{\psi} \,S\,e^{i\gamma_5\vec{\tau}\cdot\hat{r} F(r)}\,S^{\dag}\,
\psi\right]
+ \frac{{\cal I}}{4}\Tr({S}^{\dag}i\partial_t S)^2\label{gfa1}
\ee
where $\psi$ is the confined light-quark field coupled to the rotating
soliton field at the bag surface. This is the analog to Eq.(\ref{li})
(plus the rotator kinetic energy term).  Doing the
field redefinition $\psi\rightarrow S\psi$, we have
\be
L^{CB}= \int d^3x \left[\overline{\psi}
i\gamma^{\mu}\partial_{\mu}\psi
+{\psi}^{\dag}\,{S}^{\dag}i\partial_t S \,\psi\,
-\frac 12 \Delta_s \,\,
\overline{\psi} \,e^{i\gamma_5\vec{\tau}\cdot\hat{r} F(r)} \,
\psi\right]
+ \frac{{\cal I}}{4} \Tr({S}^{\dag}i\partial_t S)^2. \label{ss}
\ee
This is the exact analog to Eq.(\ref{la}). So the close analogy between the
two systems, light quarks inside a bag wrapped by a soliton and a
light antiquark of the heavy meson wrapped by a soliton, is established.

We note in passing that
the same procedure could be applied to the CK scheme. The main difference
will be
the form of the induced gauge potential.  As can be seen in Eqs.(\ref{li})
and (\ref{la}),
it consists of two parts: the generic term that leads to a Berry potential
and the terms ({\it i.e.}, the second and third terms of (\ref{li}))
which, as a consequence of the profile of the bound
heavy meson shrinking to the origin, vanish as $\sim 1/m_Q$ as the mass
of the heavy quark $m_Q$ goes to infinity.

The heavy-quark effective Lagrangian written in the form of (\ref{la})
lends itself also to a close analogy to the diatomic molecular system
described by Eq.(\ref{glagran}). To see this, we write
\be
H=\theta_a (t) H_a,\ \ \ \ \overline{H}=\theta_a^\dagger (t) \overline{H}_a
\label{Hred}
\ee
where $\theta$ is the Grassmannian introduced before, with the index $a$
representing the flavor of the light antiquark $\bar{q}_H$. Now
since the parameter space of slow rotation, $S(t)$, corresponds to the group
manifold of  $SU(2)$ which is isomorphic to $S^3$, it is convenient to use
the left or right
Maurer-Cartan forms as a basis for the vielbeins (one-form notation understood)
\be
S^{\dag} idS = - \omega_a \tau_a = -v_a^c(\phi) d\phi^c \tau^a\label{s3}
\ee
where we expressed  the ``velocity" one-form $\omega$ in the basis  of the
vielbeins $v_a^c$, and $\phi$ denotes some arbitrary parametrization
of the $SU(2)$, {\it e.g.} Euler angles.
Using (\ref{Hred}) and (\ref{s3}), we readily obtain
\be
-i\int d^3x \Tr \overline{H}(\partial_t S^\dagger S) H=i\theta_b^\dagger
(-iA_i^\alpha (\phi) T^\alpha_{ba} \dot{\phi}^i)\theta_a\label{Hla}
\ee
with
\be
-g_K T_{ba}^\alpha &\equiv& \int d^3x \Tr(\overline{H}_b I_H^\alpha H_a),
\nonumber\\
A_i^\alpha &=& 2g_K v_i^\alpha (\phi)
\ee
where $T^a$ are $n \times n$-matrix representations of the generators of $S(t)$
in the $n$-degenerate $K$ space and $g_K$ is the corresponding charge.
We have used the normalization
\be
\int d^3x \Tr \overline{H}_a H_a=-1.
\ee
We also have
\be
-i\int d^3x \Tr(\overline{H}\partial_t H) &=& i\theta_a^\dagger \partial_t
\theta_a,\label{thetakin}\\
\Tr (S^\dagger i\partial_t S)^2 &=& g^{ij} (\phi) \dot{\phi}^i \dot{\phi}^j.
\label{coll}
\ee
The Lagrangian with (\ref{Hla}), (\ref{thetakin}) and (\ref{coll})
is identical in form to that of the diatomic molecule, Eq.(\ref{glagran}).

\subsection{The vanishing of the Berry potential and symmetry restoration}
\indent

To understand the structure of the relevant Berry potential, we have to
determine the ground state of the system consisting of a light antiquark
and a skyrmion on which proper physical states are to be constructed
in the collective coordinate quantization scheme.  The interaction potential
which induces a bound state is the third term in Eq. (\ref{li}), which in
the heavy meson rest frame is\cite{foot2}
\be
H_I=-\frac{gF'(0)}{2}\int d^3 x \Tr\overline{H}H\sigma^j \tau^j.\label{hi1}
\ee
This can be rewritten in terms of the light anti-quark spin operator $S_l$
and the heavy-meson isospin operator $I_H$ as \cite{foot3}
\be
H_I=2gF'(0)\int d^3 x \Tr\overline{H}\vec{I}_H \cdot \vec{S}_l H\label{hi3}
\ee
For the $H$ classified by the $K$-spin, $\vec{K}= \vec{I}_H +\vec{S}_l$, with
the normalization $\int d^3 x \Tr\overline{H}H=-1$, we have
\be
H_I=-gF'(0)(K^2 - \frac{3}{2}).\label{hi4}
\ee
Therefore $K=0$ is a bound state for $g >0$ and $F'(0) < 0$. The $K=1$ states
are not bound unless one invokes higher-dimension terms.
This is the key point in our reasoning.

Given that the relevant state has $K=0$,
the reason for the vanishing of the Berry potential in the
heavy-meson (or heavy-quark) limit can be seen immediately.
Since the $K=0$ state is bound
and all the physical baryon states are constructed after quantization on that
state, {\em the relevant Berry potential must be
calculated on the $K=0$ state}. Although the $K=0$ state is a singlet
state with
respect to the $K$-spin, the ground state consists of at least two degenerate
states because of the two spin states of the heavy quark sitting at the origin.
The quantum numbers of the ground state of a light antiquark and a
heavy quark are the
$K$-spin, the orbital angular momentum $l$ of $\bar{q}_H$,
and the heavy-quark spin $S_Q$. There is degeneracy with respect to $S_Q$.
Therefore the Berry potential of $K=0$ for an S-wave bound state, if not zero,
is  {\em nonabelian} defined on two degenerate states ,
\be
|K=0\rangle |S_Q=+1/2\rangle,\ \ \ {\rm and} \ \ \
|K=0\rangle |S_Q=-1/2\rangle  \label{kud}
\ee
where $|K=0\rangle$ has the hedgehog configuration
\be
|K=0\rangle=\frac{1}{\sqrt{2}}\left(|\bar{d}\downarrow\rangle -|\bar{u}
\uparrow\rangle \right).
\ee
However the Berry potential {\it vanishes} in the $K=0$ state for the same
reason that the $K=0$ ground state in the chiral bag has
a vanishing Berry phase \cite{lnrz0,foot4}.
If $m_Q$ is not very large, then one should calculate also the contributions
from the second and third terms of (\ref{li}) which will come in
as $1/m_Q$ corrections.

In the discussion presented up to this point, the role of heavy-meson symmetry
is not {\it apparent}. To exhibit this, we first make the connection to the
CK scheme as used in \cite{mopr,nrz2}.
In the CK picture, the spectrum of a heavy meson bound to
a soliton is classified by the isospin and the orbital angular momentum (and
the spin for vector mesons) of the mesons, $\phi$ and $\phi^{*}$,
namely the grand spin $\Lambda = I + l + S$.
The previous argument cannot be applied to this system directly.
However, since we can construct the $K=0$ state of $\bar{q}_H$
as a linear combination of both $\phi$ and
$\phi^{*}$ which are in definite $\Lambda$ states, it is not difficult
to see how the heavy mesons conspire to give a vanishing hyperfine
coefficient, thereby making, say, the baryons $\Sigma_Q$ and $\Sigma^*_Q$
(with a heavy
quark $Q$) degenerate. For the S-wave ground state with $l=0$ given in
Eq. (\ref{kud}), the relevant states with $\Lambda=1/2$ are constructed
by a direct product of $K$ of $\bar{q}_H$ and $S_Q$ of the heavy quark
which can be decomposed into scalar and vector mesons as, for instance,
\be
\Phi_{\Lambda=1/2,\Lambda_3=1/2} &\equiv & |K=0\rangle|S_Q=1/2\rangle
= \frac 12 \left(\sqrt{2}|B^{*-}(+)\rangle -
|\overline{B^{*0}}(0) \rangle + |\overline{B^0}\rangle \right),\nonumber\\
\Phi_{\Lambda=1/2,\Lambda_3=-1/2} &\equiv & |K=0\rangle|S_Q=-1/2\rangle =
\frac 12 \left(-\sqrt{2}|\overline{B^{*0}}(-)
\rangle +|B^{*-}(0) \rangle +| B^{-}\rangle \right) \nonumber\\
\label{phi}
\ee
where $B^*$ and $B$ denoting vector and pseudoscalar $B$ mesons
respectively are used explicitly.
The spin state of the vector mesons is written in parenthesis as $+, 0, -$ for
spin $1, 0, -1$ respectively. For example, $ B^{*-}(+)$ represents the $B^{*-}$
meson of spin $S_3=+1$.
In writing Eq. (\ref{phi}), the standard recoupling of spin and isospin has
been made. Now we can see the role of heavy-meson symmetry
hidden in the $K$-spin states
by replacing the $B$ fields
in Eq. (\ref{phi}) by states with definite $\Lambda$ and $\Lambda_3$,
\be
\Phi_{1/2,1/2} &=& \frac 12 \left(\sqrt{3}|\phi^*, 1/2, +1/2 \rangle +
|\phi, 1/2, +1/2 \rangle\right),\nonumber\\
\Phi_{1/2,-1/2} &=& \frac 12 \left( \sqrt{3}|\phi^*, 1/2,
-1/2 \rangle + |\phi, 1/2, -1/2 \rangle \right)
\label{lambda}
\ee
where
\be
|\phi^*, 1/2, +1/2 \rangle &=& \frac{1}{\sqrt{3}}\left(\sqrt{2}|B^{*-}(+)
\rangle -
|\overline{B^{*0}}(0) \rangle \right),\nonumber\\
|\phi, 1/2, +1/2 \rangle &=& |\overline{B^0}\rangle,
\nonumber\\
|\phi^*, 1/2, -1/2 \rangle &=& \frac{1}{\sqrt{3}}
\left(|B^{*-}(0)\rangle -\sqrt{2}|\overline{B^{*0}}(-)\rangle \right),
\nonumber\\
|\phi^*, 1/2, -1/2 \rangle &=& | B^{-}\rangle.
\label{phib}
\ee
The conventions $|\phi^*, \Lambda, \Lambda_3 \rangle$ and $|\phi^*, \Lambda,
\Lambda_3 \rangle$ are used in the above equations.
If only the scalar meson is included, there is no way to
construct a proper bound state of $K=0$ and hence $c$ cannot vanish.
One can verify this easily by calculating the matrix element of, say, $\tau_3$
\be
\langle\phi^*, 1/2, \Lambda_3|\tau_3 |\phi^*, 1/2,
\Lambda_3 \rangle &=& -\frac 13 \Lambda_3, \nonumber\\
\langle \phi, 1/2, \Lambda_3 |\tau_3 |\phi, 1/2,
\Lambda_3 \rangle &=& \Lambda_3 .\label{tauphi}
\ee
Using  Eq. (\ref{tauphi}), we can see that the vanishing of matrix element of
$\tau_3$ for the ground state with $K=0$  is due to the {\it exact
cancellation} of the contributions from $\phi$ and $\phi^*$,
\be
\langle \Phi_{1/2,\Lambda_3}|\tau_3 \,\,| \Phi_{1/2,\Lambda_3}\rangle
 &=& \frac 14 \left( 3 \langle\phi^*,1/2, \Lambda_3 |\tau_3 |\phi^*, 1/2,
\Lambda_3 \rangle
+ \langle \phi, 1/2, \Lambda_3 |\tau_3 |\phi,1/2, \Lambda_3 \rangle
\right)\nonumber\\
 &=& 0\label{kzero}
\ee
which has been  obtained in \cite{mopr,nrz2} in a different way.
It is clear that both $\phi$ and $\phi^*$ are needed to cause
the Berry potential to vanish for a soliton-heavy meson bound state.

The disappearance of Berry potentials naturally takes place when a symmetry is
restored with a given set of states in a certain limit. In the diatomic
molecular system discussed by Zygelman\cite{zg} and reanalyzed above, a Berry
potential is obtained by slow rotation of the internuclear radius $\vec{R}$
upon
$\Sigma$ and $\pi$ states.  For finite internuclear distances $R$, these states
are not degenerate and cannot form a representation of the rotation group,
hence developing a nonvanishing
Berry (nonabelian) potential.  In the limit $R \rightarrow \infty$, however,
they become degenerate as $\kappa\rightarrow 1$
and form a representation of the rotation group and hence
the Berry potential vanishes (or more precisely it becomes a pure gauge).
In other words, when the
rotational symmetry is restored, the Berry potential developed on the
$\Sigma$ and $\pi$ states vanishes (or become pure gauge).

In the soliton-heavy meson bound system that we are considering, a
similar reasoning can be made for the vanishing Berry potential.
In the way formulated in this paper, the Berry potential vanishes because
the relevant state has $K=0$. The states with $K\neq 0$ have a nonvanishing
Berry potential. It is in {\it making} the relevant state have $K=0$ that
the heavy-quark symmetry comes in. This symmetry is not ``visible" while
the electronic rotational symmetry of diatomic molecules is, so the way
the symmetry is restored in the heavy-quark system is somewhat more
subtle. Nonetheless the mechanism is quite similar.
\section{Conclusion}
\indent

When one considers the light antiquarks in heavy mesons as fast
variables while the
heavy quark sits at the origin as a spectator,  the generic form of
Berry potentials emerges as the soliton is slowly rotated.
The bound state is composed of an antiquark and a skyrmion, characterized
by the $K$ spin, $K=0$.
The Berry potential that develops on the $K=0$ bound state is shown to vanish.
This phenomenon can be understood as the cancellation of the contributions
from scalar and vector mesons to the Berry potential or equivalently to the
hyperfine coefficient $c$ that figures in the CK picture\cite{mopr,nrz2}.
It is argued that this
cancellation is a consequence of the heavy-quark or heavy-meson
symmetry which emerges in the heavy-quark limit and that the limiting
behavior leading to a vanishing Berry potential
is quite generic as discussed in the
context of the restoration of an underlying symmetry in the dynamics.

\pagebreak

\subsubsection*{Acknowledgments}

We are grateful for valuable discussions with D.-P. Min, M.A. Nowak and
I. Zahed. HKL acknowledges the hospitality of Service de Physique
Th\'{e}orique, CE-Saclay where part of this work was done.
\vskip 2cm

\subsubsection*{Appendix}
\renewcommand\theequation{A.\arabic{equation}}
\setcounter{equation}{0}
\indent

The sign of the interaction Hamiltonian Eq.(\ref{hi4}) differs from that
of Ref.\cite{manohar}. The reason for this is sketched in this appendix.

As done in Ref.\cite{manohar}, the isospin operator $I_H^k $ on the $H$ field
is identified as
\be
I_H^k H = H \left( -\frac{\tau^k}{2} \right). \label{taus}
\ee
The action of the spin operator for the light antiquark $S_l^k$ is
more subtle. To see how it comes out, we define operationally
\be
\vec{S_l} H(P^*, P) = H( \vec{S_l} P^*, \vec{S_l} P).\label{dsl}
\ee
Let us consider for instance the action of $S_l^3$ on $H$.
For this, we write explicitly the spin wave functions for $P$ and $P^*$
in terms of the heavy-quark spin labeled by $Q$ and the light anti-quark spin
labeled by $l$
\be
P &=& \frac{1}{\sqrt{2}} \left( |\uparrow\rangle_{Q} |\downarrow\rangle_l
-|\downarrow\rangle_{Q} |\uparrow\rangle_l \right) ,\nonumber\\
P_{+}^* &=& |\uparrow\rangle_{Q} |\uparrow\rangle_l,\nonumber\\
P_{0}^* &=& \frac{1}{\sqrt{2}} \left( |\uparrow\rangle_{Q} |\downarrow\rangle_l
+|\downarrow\rangle_{Q} |\uparrow\rangle_l \right),\nonumber\\
P_{-}^* &=& |\downarrow\rangle_{Q} |\downarrow\rangle_l.\label{phl3}
\ee
Then the operation of $S_l^3$ yields
\be
S_l^3 P_{+}^* &=& \frac 12 P_{+}^*,\nonumber\\
S_l^3 P_{0}^* &=& -\frac{1}{2} P,\nonumber\\
S_l^3 P_{-}^* &=& -\frac 12 P_{-}^*,\nonumber\\
S_l^3 P &=& -\frac{1}{2}P_{0}^*.\label{sl3}
\ee
Then Eq. (\ref{dsl}) for $S_l^3$ can be rewritten
\be
S_l^3 H =\frac{1+\gamma_0}{2}\frac{1}{2}\left(P_{+}^*\gamma^{-}
+ P_{0}^*\gamma^5 -P_{-}^*\gamma^{+} -P\gamma^3 \right)\label{sl3h}
\ee
where
\be
H=\frac{1+\gamma_0}{2}\left(P_{+}^*\gamma^{-}+
P_{0}^*\gamma^3 +P_{-}^*\gamma^{+} -P \gamma^5\right)
\ee
has been used. Therefore one gets after a short algebra
\be
S_l^3 H(P^*, P)
= H(+\frac{\sigma^3}{2})\label{slh}
\ee
or more generally
\be
\vec{S}_l=H\frac{\vec{\sigma}}{2}.
\ee
Similarly the heavy quark spin, $S_Q$, acting on $H$ gives
\be
\vec{S}_Q H =  -\frac{\vec{\sigma}}{2} H.\label{sqh}
\ee
Thus the sign difference between Ref.\cite{manohar} and this paper is in
(\ref{slh}) for the identification of the light-antiquark spin operator $S_l$.


\begin{thebibliography}{99}
\bibitem{foot1}
In the CK scheme, the relevant quantum numbers are the isospin $I$ and the
angular momentum $l$ of the kaon, not those of the strange quark. The grand
spin is defined as $\Lambda = I + l$.
\bibitem{foot2}
As pointed out in \cite{mopr}, we
also have the higher-dimension term $\sim \overline{H}v^\mu H B_\mu$ where
$B_\mu$ is the topological baryon current which can induce binding.
However addition of this term does not affect our arguments.
\bibitem{foot3}
We differ in sign
from Ref.\cite{manohar}. The reason for this difference is explained in
Appendix. With this sign, we are consistent with $F(0)=\pi$ required
for our definition of the baryon current.
\end{thebibliography}
\end{document}